\documentstyle[aps,prb,manuscript]{revtex}

\begin{document}

\draft

\title{The $U=\infty$ Hubbard model with few holes: Monte Carlo
studies near half filling at nonzero temperatures.}  

\author{P\'eter~Gurin${}^{\rm 1}$ and Zsolt~Gul\'{a}csi${}^{\rm 2}$} 

\address{
${}^{\rm 1}$
Institute of Nuclear Research of Hungarian Academy of Sciences, 
POB 51, H-4001 Debrecen, Hungary.\\
${}^{\rm 2}$ 
Department of Theoretical Physics, Debrecen University,
H-4010 Debrecen, Hungary.}

\date{November, 2000}
 
\maketitle

\begin{abstract} 
We present an efficient Monte Carlo method to study the properties of the
infinitely repulsive Hubbard model with few holes at finite nonzero
temperatures. Using this method, magnetic
and specific heat results are presented for a $10 \times 10$ lattice with two
holes. Our results show that the magnetization increases with decreasing
temperature for such a finite system.
\end{abstract}

\section{Introduction}

The magnetic behaviour of the Hubbard model in the $U \: = \: \infty$ limit is
an old problem. For this case a well-known exact theorem is 
due to Nagaoka (1966) and
Thouless (1965). It provides explicit, but highly idealized conditions under
which ferromagnetism is stable in the Hubbard model. It proves that for
infinitely large repulsive Hubbard $U$ the macroscopic degeneracy of the
ground state at half filling (when the number of electrons $N$ is exactly
equal with the number of lattice sites, hereafter denoted by $N_\Lambda$) is
lifted by a single hole, i.e., when $N \: = \: N_\Lambda -1$. In this case the
saturated ferromagnetic ground state is stable for any value of the hopping
amplitude
$t$ on square, simple cubic and bcc lattices, and for $t<0$ on triangle, fcc
and hcp lattices. A very elegant and fully general proof was given for this
result by Tasaki
(1989) based on the Perron--Frobenius theorem. For the Nagaoka mechanism to
work, the lattice needs to contain loops along which the hole can move. Once
the hole moves, the maximal overlap between the initial and the final state
clearly occurs in a ferromagnetic configuration. The problems are that 
Nagaoka's
proof does not even extend to two holes, that a single hole is
thermodynamically irrelevant, and that the limit of $U=\infty$ is highly
unrealistic. Moreover, several theoretical results state the instability of
Nagaoka's ferromagnetism for two holes, see e.g. Doucot and Wen (1989),
S\"ut\H o (1991 a), T\'oth (1991).

In a recent published paper (Gurin and Gul\'acsi, 2000) a formalism was 
presented
which helps to deduce the partition function and some physical quantities
of the system based on a loop summation (i.e. path integral on lattice)
technique. The presented formulas have the advantage that the trace over the
spin degrees of freedom has been taken exactly into account. 
Resembling technique was used
by S\"ut\H o (1991 b) to analyze the magnetic behaviour of the system in
thermodynamic limit. However, his result is valid in thermodynamic limit,
only in the case $ N_h \: / \: N_\Lambda \: \rightarrow \: 0$ where $N_h
\: = \: N_\Lambda \: - \: N$ and not for finite hole concentration and $T \:
\rightarrow \: 0$.  In this paper we investigate the behaviour of a small
system with finite hole concentration, using Monte Carlo method at $T\ne 0$. 
Starting from the aim to enhance the study of magnetic properties in the
presence of a small hole concentration even at finite nonzero temperatures,
the procedure is described in detail. The studied system is a $10 \times 10$ 
lattice with periodic boundary conditions containing two holes. The results 
show an increase of the magnetisation with the decrease of the temperature. 

The remaining part of the paper is organized as follows: in Sec. II. we 
describe the representation of the Hilbert space in which the calculation is 
done, Sec.III. presents the partition function and the analyzed physical 
quantities, and Sec. IV. contains the Monte Carlo results obtained, together 
with a discussion of the deduced results.

\section{Representation of the Hilbert space}

Our Hamiltonian describes the one-band Hubbard model in 
$U \: = \: \infty$ limit:
 \begin{eqnarray}
 {\hat H}_{\infty} \: = \:
 \sum_{i,j} \: t_{i j} \: {\hat H}_{\infty}^{i j} \: = \:
 \sum_{i,j} t_{i j} \: {\hat P} \: \Bigr( \: \sum_{\sigma} \: 
          {\hat c}_{i, \sigma}^{\dagger} \: {\hat c}_{j, \sigma} \:
 \Bigl) \: {\hat P} \: , 
 \label{H} 
 \end{eqnarray} 
where $t_{i j} \: = \: - \: t$ are hopping matrix elements for nearest
neighbour sites (otherwise $t_{i j} \: = \: 0$). The operators ${\hat c}_{i,
\sigma}^{\dagger} \: ( \: {\hat c}_{i, \sigma} \: )$ creates (annihilates) an
electron with spin $\sigma$ at lattice site $i$, and ${\hat n}_{i} \: = \:
\sum_{\sigma} \: {\hat n}_{i, \sigma}$ represents the particle number
operator. The double occupancy is projected out by ${\hat P} \: = \: \sum_i \:
( \: 1 \: - \: {\hat n}_{i \uparrow} \: {\hat n}_{i \downarrow} \: )$. 

In our study we need operations with permutations therefore we fix below some
notations concerning them. We treat a permutation $\cal P$ of degree $n$ as a
bijective function of the set of the first $n$ natural numbers into
themselves. The product of the permutations is the standard function
composition: $( \: {\cal P \: Q} \:)( \: i \: ) \: = \: {\cal P}( \: {\cal Q}(
\: i \: ) \: )$. The symmetric group of degree $n$ formed by all the
permutations of degree $n$ is denoted by $S_n$. The complex Hilbert space
generated by the elements of $S_n$ as an abstract orthonormalized basis
endowed by the product defined by the convolution is called the group algebra
of the symmetric group. The parity of a permutation is $| \: {\cal P} \: |$.

We study the one band $U \: = \: \infty$ Hubbard model. 
In such a tight-bindind description the electrons can live only on 
lattice sites in two orthogonal states with spin up and down. 
Due to the kinetic energy term, an electron hops from one site
to an other (while its spin state remains unchanged), changing however
the locations of the empty and occupied sites. Hereafter, fixed positions of 
the occupied sites we call charge configuration which does not give account 
of the electronic spin states,
only their places. Iterating the effect of the hopping term
on a state several times, hopping around the lattice, the electrons produce
different charge configurations, particulary they can produce the original 
configuration again as well. 
However, the spin up and spin down electrons are commingled, therefore
the spin configuration can be different than in the original state. In our
descriptiom a spin configuration is given by the position of sites 
occupied by spin up and separately spin down electrons, respectively. 
In this way, fixing the charge
configuration and the third component of the total spin $S^z$, the spin
configuration can be described by an ordered set of the up and down spin
states: $( \: \sigma_i \: )_{i=1}^N$ in which $\sigma_i \: \in \: \{ \:
\uparrow , \: \downarrow \: \}$ indicates the spin state of the electron
situated in the $i${\em-th occupied} lattice site. Certainly, at first, the
occupied sites have to be labelled from 1 to $N$. In this way, every spin
configuration can be obtained by a permutation from 
$( \: \sigma^0 ( \: i \: ) \:
)_{i=1}^N \: = \: ( \: \uparrow , \: \uparrow , \dots , \uparrow , \:
\downarrow , \: \downarrow , \dots , \downarrow \: )$, where $\sigma^0 ( \: i
\: ) \: = \: \uparrow$ if $i \: = \: 1, \dots , N_{\uparrow} = N/2 + S^z$ and
$\sigma^0 ( \: i \: ) \: = \: \downarrow$ if $i \: = \: N_{\uparrow} + 1 ,
\dots , N$.

Based on the above mentioned observations, the Hilbert space of the system 
and a basis of it can be
expressed in terms of permutations.  The charge configuration can be described
by a function ${\cal R} \: : \: \{ \: 1 \: , \: ... \: , \: N_{\Lambda} \: \}
\: \rightarrow \: \{ \: 1 \: , \: ... \: , \: N_{\Lambda} \: \}$ for which
${\cal R}(1), \: ... \: , \: {\cal R}(N)$ are the occupied lattice sites
(taken into account an arbitrary but fixed numbering of sites), and ${\cal
R}(N+1), \: ... \: , \: {\cal R}(N_{\Lambda})$ are the positions of
holes. Actually ${\cal R}\in S_{N_{\Lambda}}$, but hereafter we denote the
charge configuration by $\tilde{\cal R}$ for which we require that
$\tilde{\cal R}(1), \: ... \: , \: \tilde{\cal R}(N)$ have to be increasingly
ordered. Moreover, we do not distingwish permutations which differ only in the
order of the positions of the holes $\tilde{\cal R}(N+1), \: ... \: , \:
\tilde{\cal R}(N_{\Lambda})$, everyone is denoted by $\tilde{\cal R}$, and the
further definitions and expressions do not depend on this undetermination. The
Hilbert space generated by all the different charge configurations as an
orthonormalized basis is denoted by ${\cal H}_c$.

To label the occupied sites from 1 to $N$ we use the function ${\tilde{\cal
R}}^{-1}$. After it, a spin configuration $( \: \sigma_i \: )_{i=1}^N$ can be
described by a permutation ${\cal P} \: \in \: S_N$ which rearrange the basic
order in $( \: \sigma^0 ( \: i \: ) \: )_{i=1}^N$ in such a way that $\sigma^0
( \: {\cal P}( \: i \: ) \: ) \: = \: \sigma_i $. This means that in a state
described by a charge configuration ${\tilde{\cal R}}$ and a spin
configuration ${\cal P}$, the electron situated in the lattice site $i$ is
in the spin state $\sigma^0 ( \: {\cal P}{\tilde{\cal R}}^{-1} ( \: i \: ) \:
)$. More precisely, there is no unique connection between the spin
configurations and permutations. A permutation uniquely determines a spin
configuration, but the reverse is not true. 
Let ${\cal Q} \: \in \: S_N$ be a permutation
which rearranges the elements of the subset $(\sigma^0)^{-1}( \: \uparrow \:)
\: = \: \{ \: 1, \: \dots \; , \: N_{\uparrow} \: \}$, and rearranges the
elements of $(\sigma^0)^{-1}( \: \downarrow \: ) \: = \: \{ \: N_{\uparrow} +
1, \: \dots \; , \: N \: \}$, but does not interchange the elements of these
two subsets. It is clear for arbitrary permutation $\cal P$, that $\cal P$ and
$\cal Q \: P$ describe the same spin configuration, because $\sigma^0 ( \:
{\cal P}( \: i \: ) \: ) \: = \: \sigma^0 ( \: {\cal Q \: P}( \: i \: ) \: )$
for all $i$. Mathematically this means the following. The above defined
permutations $\cal Q$ form a subgroup $K_{S^z} \: \subset \: S_N$, and a right
coset $\bar{{\cal P}} \: := \: \{ \: {\cal Q \: P} \: : \: {\cal Q} \: \in \:
K_{S^z} \: \}$ is in one-to-one correspondence with a spin
configuration. Moreover, the Hilbert space generated by these right
cosets as an orthonormalized basis, is isomorphic with a right ideal of the
group algebra, which will be denoted by ${\cal H}_s^{S^z}$. On this
line, an arbitrary state with given $S^z$, can be written as $\varphi_c \:
\otimes \: \varphi_s^{S^z}$ where $\varphi_c \: \in \: {\cal H}_c$ and
$\varphi_s^{S^z} \: \in \: {\cal H}_s^{S^z}$. Furthermore, $\{ \: | \:
{\tilde{\cal R}} \: \rangle \; \}$ is a basis of ${\cal H}_c$, and $\{ \: | \:
\bar{{\cal P}} \: \rangle \; \}$ is a basis of ${\cal H}_s^{S^z}$,
respectively. The
explicit expression of the above defined basis with the usual creation
operators is
 \begin{eqnarray}
 | \: {\tilde{\cal R}} \: \rangle \: \otimes \: | \: \bar{{\cal P}} \: \rangle\: 
 = \: \prod_{i=1}^{N} \: 
 {\hat c}^{\dagger}_{{\tilde{\cal R}}(i),\sigma({\cal P}(i))} \: | \: 0 \: \rangle \: ,
 \label{basis}
 \end{eqnarray} 
where $\cal P$ is an arbitrary permutation of the right coset $\bar{\cal
P}$. Now it can be seen that the requirement that $\tilde{\cal
R}(1),...,\tilde{\cal R}(N)$ have to be increasingly ordered is needed to fix
the order of the creation operators in Eq.(\ref{basis}). Let $\cal C$ a
permutation which rearranges the first $N$ numbers and leaves unchanged the
last $N_h$ ones. Using ${\cal R}^\prime \: = \: \tilde{\cal R} \: {\cal C}$
instead of $\tilde{\cal R}$ in Eq.(\ref{basis}), we get the same operators but
in different order, supposing that it is also ${\cal R}^{\prime}$ which is
used to label the occupied sites from 1 to $N$. This rearrange of the creation
operators involves a sign $(-1)^{|{\cal C}|}$. To avoid this sign ambiguity,
we use $\tilde{\cal R}$ in Eq.(\ref{basis}), for which the order of the
operators is well determined due to the requirement for the order of
$\tilde{\cal R}(1),...,\tilde{\cal R}(N)$.

\section{Partition function and physical quantities} 

Now we study the changing of a state vector from Eq.(\ref{basis}) during the
motion of the particles. Let's study the hopping of particles
step-by-step. Suppose, that the starting point is $\: {\tilde{\cal R}} \:
\otimes \: \bar{{\cal P}}$ and in the first step an electron hops from site
$i_1$ to site $j_1$, in the next step an other electron (or the same) from
$i_2$ to $j_2$ and so on. The path in the charge configuration space $\gamma
\: = \: ( \: \tilde{\cal R} \: , \: \tilde{\cal R}_1 \: , \: \dots \: , \:
\tilde{\cal R}_{l-1} \: , \: \tilde{\cal R}_l)$ is uniquely determined by the
sequence of nearest neighbour pairs $\langle \: i_1 \: , \: j_1 \: \rangle \:
, \: \dots \: , \: \langle \: i_l \: , \: j_l \: \rangle$. In the $n$th step a
hole and an electron interchange their positions. This can be described by the
transposition ${\cal P}^{i_n,j_n}$. Now we suppose, that after $l$ steps the
system gets back to the original charge configuration. However the positions
of holes are the same as in the original situation, the moving rearranged the
electrons. The electron situated originally at site $i$ moved to site $j$. The
new situation can be described by a permutation ${\cal R}_\gamma \:
\tilde{\cal R}$ where ${\cal R}_\gamma \: =\: {\cal P}^{i_l,j_l} \: \dots
{\cal P}^{i_1,j_1} \: \in \: S_{N_\Lambda}$ which takes $i$ to $j$ for all $i$
which label occupied sites and permutes somehow the labels of empty
sites. Because the spin state of the electron remains unchanged during the
hopping, the electron situated after the motion in site $j \: = \: {\cal
R}_\gamma( \: i \: )$ is in the spin state $\sigma^{\prime}_j \: = \: \sigma_i
\: = \: \sigma^0(\: \bar{\cal P} \: \tilde{\cal R}^{-1}( \: i \: ) \: ) \: =
\: \sigma^0( \: \bar{\cal P} \: \tilde{\cal R}^{-1} \: {\cal R}_\gamma ( \: j
\: ) \: )$. The charge configuration is the same as the starting ${\tilde{\cal
R}}$ therefore to recognize the final spin state ${\bar{\cal P}}^{\prime}$ we
have to write the final spin state of the electon situated in site $j$ as
$\sigma^0( \: \bar{\cal P}^{\prime} \: \tilde{\cal R}^{-1}( \: j \: ) \:
)$. As a consequence, we have $\bar{\cal P}^{\prime} \: = \: \bar{\cal P} \:
{\tilde{\cal R}}^{-1} \: {\cal R}_\gamma \: {\tilde{\cal R}} \: = \: \bar{\cal
P} \: {\cal P}_\gamma^{-1}$ where ${\cal P}_\gamma \: = \: {\tilde{\cal
R}}^{-1} \: {\cal R}_\gamma^{-1} \: {\tilde{\cal R}}$ is an element of
$S_{N_{\Lambda}}$ but it rearranges the first $N$ numbers among themself and
leaves unchanged the last $N_h$ ones, therefore it can be considered as an
element of $S_N$. Furthermore the final charge configuration is again
$\tilde{\cal R} \: = \: ( \: {\cal R}_\gamma \: {\cal R} \: ) \: {\cal
P}_\gamma$ instead of ${\cal R}_\gamma \: \tilde{\cal R}$ therefore we have to
take into account a sign $(-1)^{|{\cal P}_\gamma |}$ --- as we mentioned at
the end of the previous Section --- to write the final state as $\:
{\tilde{\cal R}} \: \otimes \: (-1)^{|{\cal P}_\gamma |} \: \bar{{\cal P}} \:
{\cal P}_\gamma^{-1} $

The mapping $\bar{\cal P} \: \rightarrow \: (-1)^{|{\cal P}_\gamma |} \:
\bar{\cal P} \: {\cal P}_\gamma^{-1}$ is defined on the basis of ${\cal
H}_s^{S^z}$. The influence caused by a hopping of holes along a loop $\gamma$
is described by a linear operator which is the linear extension of the mapping
$\bar{\cal P} \: \rightarrow \: (-1)^{|{\cal P}_\gamma|} \: \bar{\cal P} \:
{\cal P}_\gamma^{-1}$ to the whole ${\cal H}_s^{S^z}$. It is actually a linear
representation of the symmetric group $S_N$ on ${\cal H}_s^{S^z}$ (Gurin and
Gul\'acsi, 2000) hereafter denoted by $T^{S^z}[ \: {\cal P}_\gamma \: ]$.

Based on the above considerations, we
can give expressions for the partition function as follows
\begin{eqnarray}
Z \: = {\rm Tr} \: e^{-\frac{{\hat H}_{\infty}}{k T} }
\: = \: \sum_{S^z} \: 
{\rm Tr}_{{\cal H}^{S^z}} \: e^{-\frac{{\hat H}_{\infty}}{k T} } \: ,
\label{ter2}
\end{eqnarray}
and the expectational value of the square of the total spin 
\begin{eqnarray}
\langle \: {\hat S}^2 \: \rangle \: 
= \: \frac{1}{Z} \: 
  {\rm Tr} ( \: \hat S^2 \: e^{-\frac{{\hat H}_{\infty}}{k T} } \: ) \:
= \: \frac{1}{Z} \: \sum_{S^z} \: 3 \: ( \: S^z \: )^2 \:
{\rm Tr}_{{\cal H}^{S^z}} \: e^{-\frac{{\hat H}_{\infty}}{k T} } 
\label{ter1}
\end{eqnarray}
in which the sum over the spin degrees of freedom is expressed analitically in 
terms of the characters of the symmetric group. From the above expressions
it can be seen that what we need for calculations is in fact 
\begin{eqnarray}
{\rm Tr}_{{\cal H}^{S^z}} \: e^{-\frac{{\hat H}_{\infty}}{k T} } \:
&=& \: \sum_{l=0}^\infty \: \frac{1}{l \: !} \: \left(\frac{t}{k \: T}\right)^l \: \sum_{\tilde{\cal R}}
\: \sum_{\gamma \in \Omega_{\tilde{\cal R}}(l)} \: \sum_{ \bar{\cal P} } \: 
  \langle \: \bar{\cal P} \: | \: T^{S^z}[ \: {\cal P}_{\gamma} \: ]
   \: | \: \bar{\cal P} \: \rangle \: 
\nonumber \\
&=& \: \sum_{l=0}^\infty \: \frac{1}{l \: !} \: \left(\frac{t}{k \: T}\right)^l \: \sum_{\tilde{\cal R}}
\: \sum_{\gamma \in \Omega_{\tilde{\cal R}}(l)} 
  \chi^{S^z}( \: {\cal P}_{\gamma} \: ). 
\label{ter3} 
\end{eqnarray}
where $\Omega_{\tilde{\cal R}}( \: l \: )$ denotes the set of the $l$-length
loops (closed pathes) with starting point $\tilde{\cal R}$ in the space of
charge configurations, and $\chi^{S^z}$ is the character of the
representation $T^{S^z}$. Every character is constant on an arbitrary
conjugate class $C$. Therefore, the above sum over $\Omega_{\tilde{\cal R}}(
\: l \: )$ has $N^{(C)}_{\tilde{\cal R}}( \: l \: )$ identical members, where
$N^{(C)}_{\tilde{\cal R}}( \: l \: )$ represents the number of the loops with
length $l$ and starting point ${\tilde{\cal R}}$ for which ${\cal P}_{\gamma}
\: \in \: C$.  Now we use (see Gurin and Gul\'acsi, 2000)
\begin{eqnarray}
\sum_{S^z = -N/2}^{N/2} \: \chi^{S^z}( \: C \: ) \:
= \: ( \: - \: 1 \: )^{|C|} \: 2^{\sum_{i=1}^N C_i} \: ,
\label{sum_of_chi}
\end{eqnarray}
furthermore
\begin{eqnarray}
\sum_{S^z = -N/2}^{N/2} \: ( \: S^z \: )^2 \: \chi^{S^z}( \: C \: ) \:
= \: ( \: - \: 1 \: )^{|C|} \: 2^{\sum_{i=1}^N C_i} \:
  \left( \: \frac{1}{4} \: \sum_{i=1}^N \: i ^2 \: C_i \: \right) \: .
\label{sum2_of_chi}
\end{eqnarray}
Here $| \: C \: |$ is the parity of permutations from the conjugate
class $C$ and $( \: C_1 \: , \: C_2 \: , \: \dots \: , \: C_N \: )$
describes their cycle structure, i.e. these permutations contain 
$C_i$ cycles with length $i$.

Finally, inserting Eqs.(\ref{ter3},\ref{sum_of_chi},\ref{sum2_of_chi}) into
Eqs.(\ref{ter2},\ref{ter1}) we obtain
\begin{eqnarray}
  Z \: 
= \sum_{l=0}^{\infty} \: 
  \frac{1}{l \: !} \: \left(\frac{t}{k \: T}\right)^l \:
 \sum_{\tilde{\cal R}} \:\sum_{C \subset S_N} \: 
  N^{(C)}_{\tilde{\cal R}}( \: l \: ) \: ( \: - \: 1 \: )^{|C|} \: 
  \left( \: 2^{\sum_{i=1}^N C_i} \right) \: ,
\label{Z}
\end{eqnarray} 
and
\begin{eqnarray}
  \langle \: {\hat S}^2 \: \rangle \: 
= \: \frac{3}{4 \: Z} \: \sum_{l=0}^{\infty} \: 
  \frac{1}{l \: !} \: \left(\frac{t}{k \: T}\right)^l \:
  \sum_{\tilde{\cal R}} \: \sum_{C \subset S_N} \: 
  N^{(C)}_{\tilde{\cal R}}( \: l \: ) \: ( \: - \: 1 \: )^{|C|} \: 
  \left( 2^{\sum_{i=1}^N C_i} \: \right) \:
  \left( \: \sum_{i=1}^N \: i^2 \: C_i \: \right)
 \: .  
\label{S^2}
\end{eqnarray}

To determine the coefficients $N^{(C)}_{\tilde{\cal R}}( \: l \: )$ we should
follow the following procedure.  
Starting from a charge configuration $\tilde{\cal R}$, holes moves $l$ steps 
in such a way that finally the sysyem goes back to the
original hole configuration (this is a loop with length $l$). In this process,
an arbitrary hole can be moved in every step, the order of the steps being
relevant. A step means that we interchange a hole with an electron from its
neighbourhood. Counting the number of $l$-loops for which the cycle structure
of the permutation of electrons after the $l$th step is $C$, we get the number
$N^{(C)}_{\tilde{\cal R}}( \: l \: )$.

\section{Monte Carlo results and discussions} 

The coefficients $N_{\tilde{\cal R}}^{(C)}( \: l \: )$ can be determined by
Monte Carlo method. By random sampling we can obtain the percentage of loops
with a given length for which ${\cal P}_{\gamma} \: \in \: C$. The
Eqs.(\ref{Z}) and (\ref{S^2}) have the advantage that the trace over the spin
degrees of freedom has been taken already. Hence we need only to sample the
loops (i.e. ``world lines'') of holes instead of the loops of the
electrons. This is advantageous since due to the presence of different 
possible electron spin configurations, the electron loops have $2^N$ times 
more starting points than the hole loops. 

We studied by this method a 2D $10 \times 10$ lattice 
with periodic boundary conditions
in the presence of two holes. We took into account every possible starting
hole (charge) configuration. Because of the periodic boundary conditions,
translational, rotational and reflection symmetries of the lattice,
there are 20 essentially different charge configurations instead of 
${ 100 \choose 2}$. Because of the fact that the contributions of the 
$2^{98}$ different spin configurations corresponding to a fixed charge 
configuration are taken into account analitycally by Eqs.(\ref{Z},\ref{S^2}), 
we have taken into account every possible strating point of loops. 
Furthermore, starting from a charge configuration, we took 
into consideration every path  up to  the length $l \: = \: 14$. 
Thus, we counted exactly
every loop up to $l \: \leq \: 14$, therefore the coefficients of our high
temperature expansion are exact up to 14th order in $\beta \: / \: | \: t \:
|$. Then, we continued every path with random directions up to $l \: = \:
100$. 

The results can be seen in the Fig. 1. The continuous lines show the result of
the high temperature expansion with exact coefficients up to 14th
order. Taking into account further coefficients determined by Monte Carlo
method up to 50th order we get the short-dash lines, and up to 100th order we
get the long-dash lines. The expectation value of the square of the total
spin goes to $3 \: N \: / \: 4 \: = \: 73.5$ (and not to zero) as $T$ goes to
infinity. It is understandable, because at very high temperature the
likelyhood of every possible states become equally $1\: / \: Z $. (The
spectrum of the Hamiltonian is bounded.) Therefore, the expectation value of
an arbitrary quantity goes to its simple algebraic average over the all
states. This gives the result mentioned above for $\langle \: \hat S^2 \:
\rangle$). This means in fact paramagnetic behaviour, because a smaller
total spin value has greater thermodynamic weigh, i.e. the dimension of the 
eigensubspace of $\hat S^2$ is greater, therefore the $S \: = \: 0$ subspace 
is the greatest one, so this is the most probable value of the spin. 
Because of this reason, the total spin (but not
$\langle \: \hat S^2 \: \rangle$) per particle is proportional to $1\: / \:
\sqrt{ N }$ which goes to zero in thermodynamical limit.

Because of the inaccuracy and the complete neglection of the higher order
coefficients in our expressions, the numerical error increases as the
temperature decreases. Therefore we do not plot the $T \to 0$ temperature 
region in Fig. 1. Nevertheless, in the presented region the results are
correct since the higher order terms do not improve the lower order 
contributions in this domain. At these temperatures however, due to the 
Monte Carlo method itself, small numerical errors are present of about $1\%$.

The obtained results show that the spontaneous magnetization increases as the
temperature decreases. This tendency becomes stronger when we use more
accurate approximation. At the same temperature range when $\langle \: \hat
S^2 \: \rangle$ starts to increase, the specific heat has a maximum. Hovewer
this peak in the specific heat seems to be significant --- in higher order
approximations the curves have to break down as well since the specific heat
has to go to zero in $T \to 0$ limit, --- but the increase in $\langle \:
\hat S^2 \: \rangle$ is small, nevertheless very fast.

Related to the two and few hole cases in the literature it have been published 
(Doucot and Wen 1989, S\"ut\H o 1991 a, T\'oth 1991) that the ground state can 
not have maximal spin. This statement is based on variational calculations. 
There exists trial state with spin $S \: = \: S_{max} \: - \: 1$ which have 
lower energy than the fully polarized state with spin $S_{max}$. However, up 
to now nobody can construct variational state with spin $S \: \leq \: 
S_{max} \: - \: 2$ which would have lower energy, thus these results do not 
exclude the ferromagnetic behaviour, only the highest spin value. 
Simultaneously our results suggest that the very low spin values can be exclude 
as well, and the ground state is not a singlet. As a conclusion we may state 
that the system behave at low temperature as a not fully saturated ferromagnet.
This is consistent with exact diagonalization results for small clusters 
(Arrachea 2000).

\section{Acknowledgements}

P.G. kindly acknowledge the Young Research Fellow Scholarship of the Institute
of Nuclear Research of Hungarian Academy of Sciences, Debrecen. For Zs.G.
research supported by FKFP-0471 and OTKA-022874 of Hungarian Founds for
Scientific Research.

\newpage

\section*{References}

	\mbox{}\\
{\sc Arrachea, L.,} 
	2000, {\sl Phys. Rev. B,} {\bf 62}, 10033.
\\
{\sc Doucot, B.,} and {\sc Wen, X. G.,}
	1989, {\sl Phys. Rev. B,} {\bf 40}, 2719.
\\
{\sc Gurin, P.,} and {\sc Gul\'acsi, Zs.,}
	{\sl Phil. Mag. B,} 2000, in press
\\
{\sc Nagaoka, Y.,} 
	1966, {\sl Phys. Rev.} {\bf 147}, 392. 
\\
{\sc S\"ut\H o, A.,} 
	1991 a, {\sl Commun. Math. Phys.,} {\bf 140}, 43;
	1991 b, {\sl Phys. Rev. B,} {\bf 43}, 8779.
\\
{\sc Thouless, D. J.,}
	1965, {\sl Proc. Phys. Soc.,} {\bf 86}, 893.
\\
{\sc Tasaki, H.,} 
	1989, {\sl Phys. Rev. B,} {\bf 40}, 9192.
\\
{\sc T\'oth, B.,} 
	1991, {\sl Lett. in Math. Phys.,} {\bf 22}, 321.

\begin{figure} 
\caption{ 
(a): square of the total spin and (b): specific heat as the function of
the temperature measured in $| \: t \: |$ units. The continuous (lower)
lines show the result of a 14th order high temperature expansion, the short-dash
lines (in the middle) show the result of an 50th order high temperature
expansion, the long-dash (upper) lines of a 100th order expansion based on
coefficients determined by Monte carlo method.}
\end{figure}

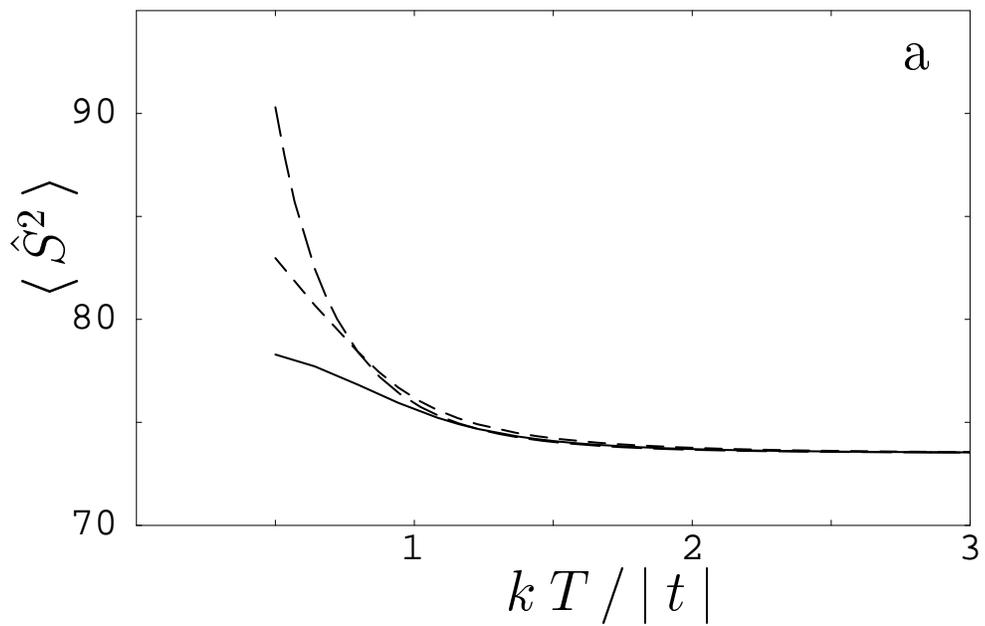
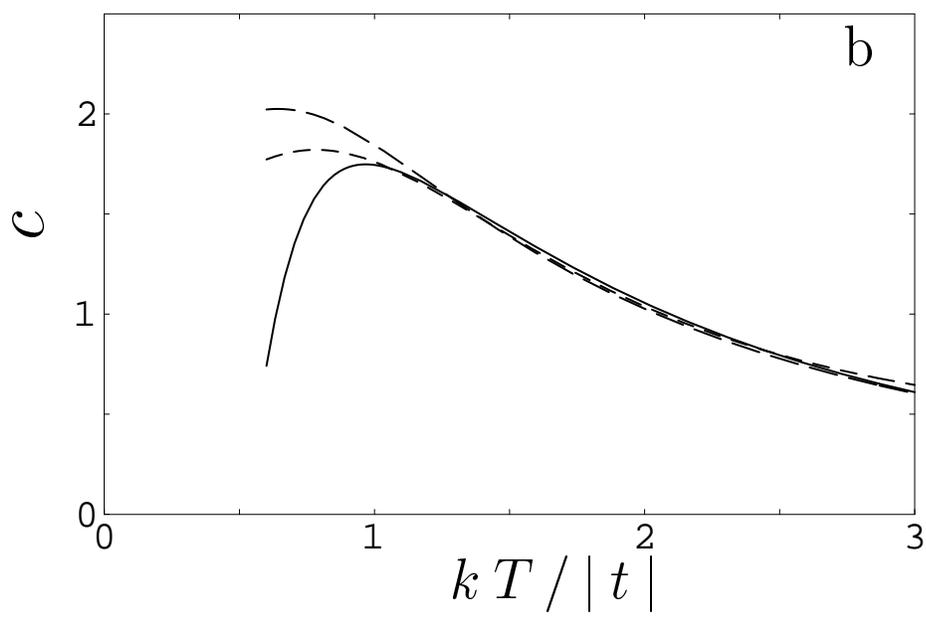
\end{document}